\documentclass[12pt]{iopart}
\usepackage{placeins}
\usepackage {epsfig} \usepackage {xspace} \expandafter\let\csname
equation*\endcsname\relax \expandafter\let\csname
endequation*\endcsname\relax \usepackage {amsmath} \usepackage
{graphicx} \usepackage [english] {babel} \usepackage {amsfonts}
\usepackage {amssymb} \usepackage {latexsym} \usepackage {color}
\usepackage {ulem} \usepackage {iopams}
\newcommand{\op}[1]{\hat{\rm #1}} \newcommand{\ket}[1]{\left| #1
  \right>} 
\newcommand{\braket}[2]{\left< #1 \vphantom{#2} \right|
  \left. #2 \vphantom{#1} \right>} 
\newcommand{\QMa}[3]{\left< #1 \vphantom{#2#3} \right| #2 \left| #3
  \vphantom{#1#2} \right>} 
\newcommand{\eqn}[1]{\begin{equation}
 			\begin{split}
                       #1
 			\end{split}
                      \end{equation}}

\newcommand{\figref}[1]{Fig.~\ref{#1}}

\newcommand{\secref}[1]{Section~\ref{#1}}

\renewcommand{\eqref}[1]{Eq.~(\ref{#1})}
\newcommand{\tabref}[1]{Tab.~\ref{#1}}

\begin{document}

\title{Resonances in ultracold dipolar atomic and molecular gases}

\author{Bruno Schulz, Simon Sala, and Alejandro Saenz}

\address{AG Moderne Optik, Institut f\"ur Physik,
  Humboldt-Universit\"at zu Berlin, Newtonstrasse 15, 12489 Berlin,
  Germany}
\ead{alejandro.saenz@physik.hu-berlin.de}
\vspace{10pt}
\begin{indented}
\item[]\today
\end{indented}


\begin{abstract}
  A previously developed approach for the numerical treatment of two 
  particles that are confined in a finite optical-lattice potential 
  and interact via an arbitrary isotropic interaction potential 
  has been extended to incorporate an additional anisotropic 
  dipole-dipole interaction. The interplay of a model but realistic 
  short-range Born-Oppenheimer potential and the dipole-dipole interaction 
  for two confined particles is investigated. A variation of the 
  strength of the dipole-dipole interaction leads to diverse resonance 
  phenomena. In a harmonic confinement potential some resonances show 
  similarities to $s$-wave scattering resonances while in an anharmonic 
  trapping potential like the one of an optical lattice inelastic 
  confinement-induced dipolar resonances occur. The latter are due to  
  a coupling of the relative and center-of-mass motion caused by the 
  anharmonicity of the external confinement.
\end{abstract}


\section{Introduction}

In recent years significant experimental progress has lead to
sophisticated cooling and trapping techniques of polar molecules and
of atomic species having a large dipole moment
\cite{dipol:stwa04,dipol:li05,dipol:bara12,dipol:laha09}. A very promising approach for achieving
ultracold polar molecules is the formation of weakly-bound molecules
making use of a magnetic Feshbach resonance and a subsequent transfer to
the ground state using the STIRAP (stimulated Raman adiabatic passage) 
scheme \cite{dipol:ospe08}. Based on this method gases of
motionally ultracold RbK \cite{dipol:ni08} or LiCs
\cite{dipole:deig08} molecules in their rovibrational ground
states were achieved. This fascinating progress paved the way towards
degenerate quantum gases with predominant dipole-dipole interactions
(DDI). In the case of magnetic dipoles the Bose-Einstein condensation
(BEC) of $^{52}\text{Cr}$, an atom with a large magnetic moment of 6
$\mu_B$, was already achieved in 2004 \cite{dipol:stuh05}. Although
in chromium the DDI can be enhanced relative to the atomic short-range
interaction by decreasing the strength of the latter using a Feshbach
resonance \cite{dipol:grie05}, the DDI is typically still
smaller or at most of the same magnitude as the van-der-Waals forces. In
order to create an atomic gas with a DDI larger than the van-der-Waals
forces, it was possible to realise a BEC of Dysprosium
($10\mu_B$) \cite{dipol:lu11} and Erbium ($12\mu_B$)
\cite{dipol:aika12}.

The properties of the DDI are completely different from those of
isotropic short-range interactions, e.\,g., the ones between two atoms 
with no DDI. The
DDI has a \textit{long-range} character as it decays as $1/r^3$, where
$r$ is the inter-particle distance, and it is \textit{anisotropic} which
means that even the sign of the interaction depends on the angle
$\theta$ between the polarisation direction and the relative
position of the particles. 

A full and quantitative 
understanding of the behaviour of two particles in an external 
trapping potential is a prerequisite for the manipulation and control of
ultracold two-body systems that have been proposed for possible 
quantum-computer realisations \cite{dipol:demi02,cold:schn12}. This  
knowledge about the microscopic two-body physics is also important 
for the understanding of the rich physics of the corresponding many-body 
systems. For example, more accurate Bose-Hubbard parameters for describing 
an ultracold quantum gas in an optical lattice have been extracted from two-particle
calculations in the absence \cite{cold:schn09} or the presence of 
DDI \cite{dipol:wall13}. Furthermore, inelastic confinement-induced 
two-body resonances \cite{cold:sala12,cold:sala13} were found to have caused 
massive atom losses also in quantum gases with many particles \cite{cold:hall10b}. 
Evidently, the possibility to vary the interaction strength within ultracold 
atomic quantum gases with the aid of two-body magnetic Feshbach resonances 
has been of paramount importance for the whole research field. 

Different theoretical approaches to describe dipolar systems have been
reported earlier. To describe an ultracold dipolar gas
trapped in a harmonic confinement, in
\cite{dipol:yi00,dipol:yi01} a pseudo potential as a
function of the dipole moment is proposed for the short-range interaction, 
while for the long-range potential the anisotropic DDI is
adopted. The approach described in \cite{dipol:shi12} uses a
hard-sphere potential for the short-range part of the interaction potential between two dipolar
particles in free space. In these approaches $s$-wave like 
scattering resonances induced by the
dipolar interactions were observed. Following the prediction of elastic
confinement-induced resonances in quasi-1D confinement
\cite{cold:olsh98,cold:berg03}, such resonances for dipolar
systems were considered in \cite{dipol:sinh07,dipol:shi14}. 

The approach described in the present work extends the method
described in \cite{cold:gris11} which allows for the use of a
realistic, numerically given Born-Oppenheimer potential curve for the
short-range part of the interaction potential by adding an additional
DDI. In contrast to the use of a $\delta$ pseudo-potential 
that supports a single bound state, the
use of a realistic short-range potential supports often many deeply bound
states. Furthermore, in contrast to numerous previous works, see 
e.\,g.\ \cite{dipol:rone06,dipol:bort06,dipol:sinh07,dipol:bart13}, in 
our approach the 
external trap potential is chosen as a finite optical-lattice potential,
i.\,e., a truncated Taylor series for a $\sin^2$ or $\cos^2$
optical-lattice potential. While a truncation at the second order
yields a harmonic trapping potential, anharmonic multi-well
potentials can be achieved, if the truncation is performed at higher
orders. In this work it will be demonstrated that in an anharmonic trapping
potential the coupling of center-of-mass and relative motion leads to
the occurrence of inelastic resonances at which bound states with some 
center-of-mass excitation couple to states of unbound dipoles. This
mechanism is analogous to the inelastic confinement-induced
resonances described in \cite{cold:sala12,cold:peng11,cold:sala13} for ultracold
atomic systems without DDI.

This article is organised as follows. First, the Hamiltonian and the
numerical method are introduced in \secref{ch:two} and
\secref{ch:method}, respectively. In \secref{ch:results} the 
influence of the dipolar interaction is investigated for two different 
trapping potentials. The results for an isotropic harmonic trap are 
shown in \secref{ch:harmonic}, the ones for an anisotropic sextic 
potential with coupling between center-of-mass and relative motions 
in \secref{ch:sextic}. A conclusion is provided in 
\secref{ch:summary}.

\section{Two-Body Problem of Trapped Dipolar Particles}
\label{ch:two}

A system of two dipolar particles with masses $m_1$, $m_2$ and the absolute coordinates
$\bold{r_1}$, $\bold{r_2}$ trapped in an optical lattice is
described by the Hamiltonian
\eqn{\op{H} = \op{T}_1(\bold{r_1}) + \op{T}_2(\bold{r_2})+
  \op{V}_{1}(\bold{r_1})+ \op{V}_{2}(\bold{r_2}) + \op{V}_{\rm
    int}(\bold{r_1}-\bold{r_2}) \; ,\label{eq:Hamiltonian}}
where $\op{T}_i(\bold{r_i})$ is the kinetic energy operator
$\frac{\hat{\bold{p}}_i^2}{2 m_i}$ for each particle. The trapping
potential is chosen as a $\sin^2$ optical lattice\footnote{Note, also
  $\cos^2$ lattice potentials are implemented in the code. For
  simplification of the notation, only the $\sin^2$
  lattices are explicitly considered here.}
\eqn{V_i(\bold{r_i})=\sum_{j=x_i,y_i,z_i} V_{0,j} \sin ^2(k_j j)}
which can be experimentally obtained by the superposition of
counter-propagating laser beams \cite{cold:bloc05}, where the
parameters $V_{0,j}$ and $k_j$ are the potential depth and the
components of the wavevector $k_j=2\pi / \lambda_j$, respectively.

Relative (rm) and center-of-mass (CM) motion coordinates
$\bold{r}=\bold{r_1}-\bold{r_2}$, $\bold{R}=\mu_1 \bold{r_1}+\mu_2
\bold{r_2}$, respectively, are introduced to transform the two-body
Hamiltonian in \eqref{eq:Hamiltonian} into
\eqn{\op{H}(\bold{r},\bold{R}) = \op{h}_{\rm rm}(\bold{r}) +
  \op{h}_{\rm CM}(\bold{R}) +
  \op{W}(\bold{r},\bold{R}) \label{eq:Ham_rm}}
where $\op{h}_{\rm rm}$ and $\op{h}_{\rm CM}$ are the separable parts
of the Hamiltonian in relative and center-of-mass coordinates,
respectively, and $\mu_i=\frac{m_i}{m_1+m_2}$. The coupling term
$\op{W}(\bold{r},\bold{R})$ describes the non-separable parts of the
Hamiltonian that originate from the non-separability of the
optical-lattice potential in relative and center-of-mass
coordinates. The center-of-mass motion part 
\eqn{\op{h}_{\rm CM}(\bold{R})=\frac{\hat{\bold{P}}^2}{2M} + V_{\rm
    CM}(\bold{R})}
of the Hamiltonian in \eqref{eq:Ham_rm} and the relative-motion part
\eqn{\op{h}_{\rm rm}(\bold{r})=\frac{\hat{\bold{p}}^2}{2\mu} + V_{\rm
    rm}(\bold{r}) + V_{\rm int}(\bold{r})\;, \label{eq:h_rm} }
contain the respective momentum operators $\hat{\bold{p}}$ and
$\hat{\bold{P}}$ in relative and center-of-mass coordinates, 
i.\,e.\ $\hat{\bold{p}}=\hat{\bold{p_1}}-\hat{\bold{p_2}}$ and 
$\hat{\bold{P}}=\mu_1\hat{\bold{p_1}}+\mu_2 \hat{\bold{p_2}}$. While
the implementation of the algorithm also allows for distinguishable
particles, in the present work only the special case of identical
particles is considered which is in accordance with many  
experiments \cite{dipol:stwa04,dipol:li05, dipol:aika12,dipol:lu11}. 
In this case, the separable parts of the optical-lattice potential, 
\eqn{ V_{\rm CM}(\bold{R}) = 2\sum_{c=x,y,z}V_{0,c}\sin^2 \left(k_c
  R_c\right)}
and
\eqn{ V_{\rm rm}(\bold{r}) =
  2\sum_{c=x,y,z}V_{0,c}\sin^2\left(\frac{k_c r_c}{2}\right)\;, \label{eq:v_rm} }
and the coupling part
\eqn{\op{W}(\bold{r},\bold{R})=-4 \sum_{c=x,y,z} V_{0,c}
  \sin^2\left(k_c R_c\right) \sin^2\left(\frac{k_c r_c}{2}\right) }
keep a simple form containing the $\sin^2$ terms.

The interaction potential $V_{\rm int}(\bold{r})=V_{\rm
  int}(r,\theta)$ in \eqref{eq:h_rm} describes the interaction of two dipolar
particles. In the present approach, the interaction potential consists
of an isotropic short-range part $V_{\rm sh}$ and of the long-range
DDI $V_{\rm dd}$ for aligned dipoles along the $z$ axis, 
\eqn{V_{\rm int}(r,\theta) = V_{\rm sh}(r)+ V_{\rm dd}(r,\theta) =V_{\rm sh}(r)+ \frac{C_{\rm dd}}{4\pi}
  \frac{1-3\cos^2(\theta)}{r^3} = V_{\rm sh}(r)
  -\sqrt{\frac{16\pi}{5}} \frac{C_{\rm dd}}{4\pi}
  \frac{Y_{2}^{0}(\theta)}{r^3} \;, \label{eq:v_dd}}
where $C_{\rm dd}=d_1 d_2 / \epsilon_0$ 
($C_{\rm dd}= \mu_0 \tilde{\mu}_1 \tilde{\mu}_2$) is the coupling constant for
the electric dipole moments $d_1$ and $d_2$ (magnetic dipole moments 
$\tilde{\mu}_1$ and $\tilde{\mu}_2$) where $\epsilon_0$ is the vacuum 
permittivity ($\mu_0$ the vacuum permeability). Furthermore, in \eqref{eq:v_dd} the 
spherical harmonic $Y_2^0(\theta) = \sqrt{5/(16 \pi)} (3 \cos^2(\theta)-1)$ 
is introduced. The alignment of the electric or magnetic dipoles can be
obtained with static electric or magnetic fields, respectively.
Additionally, as the applied electric field increases, the dipole
moment $d$ continuously increases from zero to the full permanent dipole
moment in the case of electric dipoles.
\begin{figure}[htpb]
\centering \includegraphics[width=\linewidth]{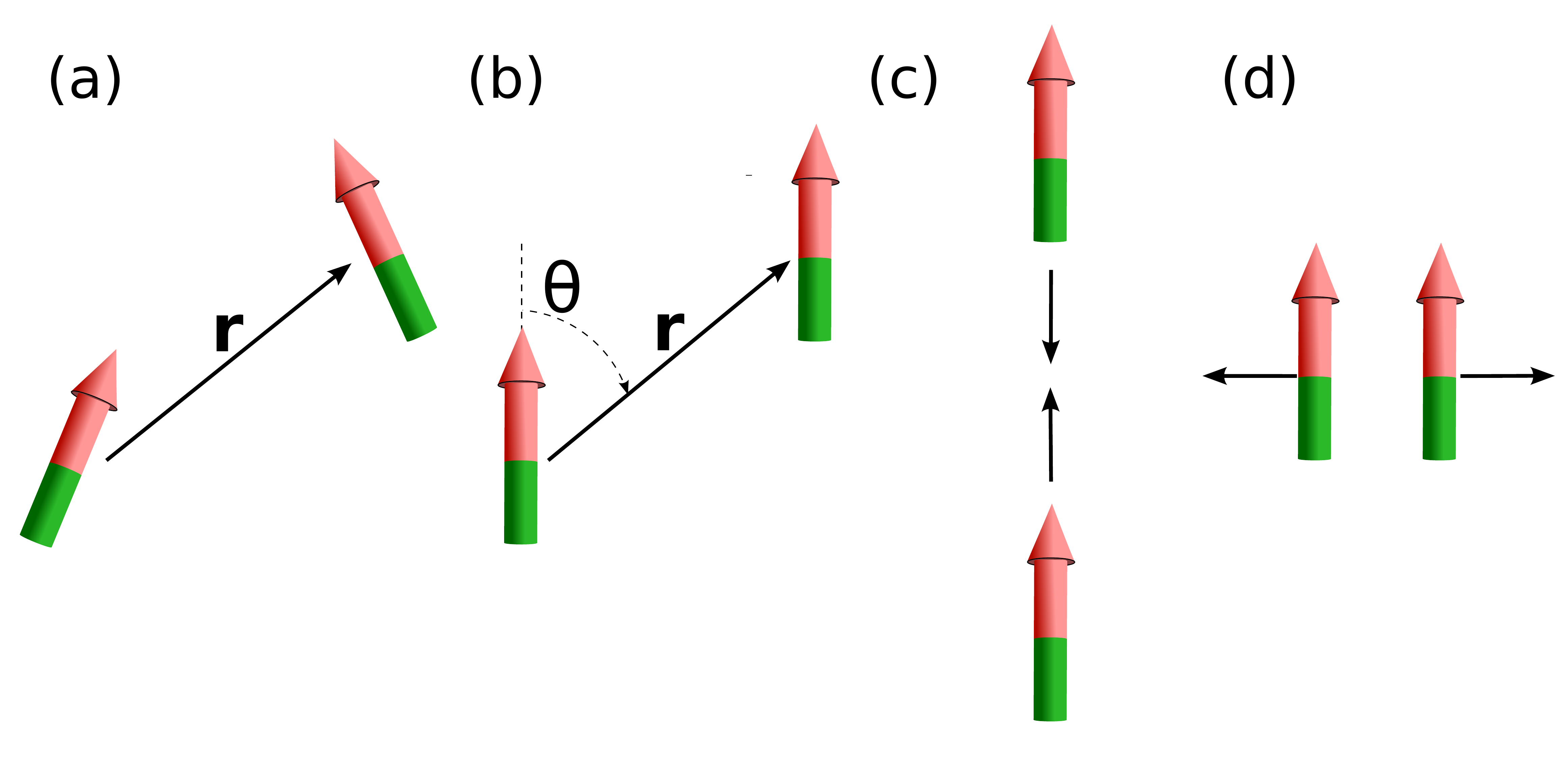}
\caption{Two particles interacting via a dipole-dipole interaction (DDI): 
  (a) non-polarised case; (b) with polarisation along the
  $z$ axis; (c) two polarised dipoles in the head-to-tail
  configuration attracting each other (black arrows); (d) two
  polarised dipoles in the side-by-side configuration repelling each
  other (black arrows).}
\label{fig:t_lahaye_bild_dipole}
\end{figure}

A rather unique feature of the present approach is that 
the short-range interaction potential $V_{\mathrm{sh}}$ can be chosen 
arbitrarily as some analytical expression, some numerically given 
potential curve, or a mixture of both. The only constraint is its 
isotropy. Choosing a realistic atomic or molecular interaction
potential provides the unique opportunity to investigate in a
realistic fashion especially the regime where both, the isotropic short-range
interaction as well as the DDI, have a comparable influence. Such a
study is not possible with model potentials (such as, e.\,g.,
zero-range potentials \cite{dipol:bart13} or hard spheres 
\cite{dipol:kanji08}) that do not realistically 
reproduce the behavior of the tail of the short-range potential.

The anisotropy of the DDI is described by the spherical harmonic
$Y_{l=2}^{m=0}(\theta)$, see \eqref{eq:v_dd}. The DDI is repulsive for
dipoles in the side-by-side configuration and attractive in the
head-to-tail configuration, see \figref{fig:t_lahaye_bild_dipole}.
Therefore, with increasing dipole moment the overall interaction
potential changes dramatically from the generic isotropic
short-range interaction to an anisotropic long-range DDI. In addition,
the shape of the long-range part of the wavefunction is strongly
influenced by the trap and a possible centrifugal barrier.

\FloatBarrier

\section{Method}
\label{ch:method}

The full treatment of two particles interacting by an arbitrary
isotropic interaction potential and confined in a finite optical
lattice has been introduced in \cite{cold:gris11}. In the present work
this approach has been extended for treating particles that interact
with an additional anisotropic DDI. Details of the original
approach with isotropic interactions can be found in \cite{cold:gris11}. 
Here, this approach is briefly described and the extension for the 
inclusion of the DDI is presented.

\subsection{Exact Diagonalisation\label{ch:diag}}

For a given trapping potential the Schr\"odinger equation
\eqn{\op{H} \ket{\Psi_i}= \mathcal{E}_i \ket{\Psi_i}}
of the Hamiltonian of \eqref{eq:Ham_rm} in relative and center-of-mass
coordinates
is solved by expanding $\Psi$ in terms of configurations $\Phi$,
\eqn{\Psi_i(\bold{R},\bold{r})=\sum_k
  \mathcal{C}_{ik}\;\Phi_k(\bold{R},\bold{r})\;.}
The configurations
\eqn{\Phi_k(\bold{R},\bold{r})=\varphi_{i_k}(\bold{r})
  \psi_{j_{_k}}(\bold{R})}
are products of the eigenfunctions $\varphi$ and $\psi$ that are the
solutions of the eigenvalue equations
\eqn{\op{h}_{\rm rm}\ket{\varphi_i}=\epsilon_i^{\rm
    rm}\ket{\varphi_i}, \qquad \op{h}_{\rm
    CM}\ket{\psi_j}=\epsilon_j^{\rm
    CM}\ket{\psi_j}\;. \label{eq:SGL_rm}}
of the Hamiltonians for the relative and center-of-mass motions,
respectively.
%
%
%
Once the eigenvectors $\ket{\varphi}$ and $\ket{\psi}$ are obtained
from the solution of the corresponding generalised matrix eigenvalue problem with the
matrices
\eqn{h_{\bold{a},\bold{a'}}^{\rm rm}=
  \QMa{\varphi_{\bold{a}}}{\op{h}_{\rm rm}}{\varphi_{\bold{a'}}},
  \qquad s_{\bold{a},\bold{a'}}^{\rm
    rm}=\braket{\varphi_{\bold{a}}}{\varphi_{\bold{a'}}}\;,\label{eq:Hamiltonian_rm}}
\eqn{h_{\bold{b},\bold{b'}}^{\rm CM}=
  \QMa{\psi_{\bold{b}}}{\op{h}_{\rm CM}}{\psi_{\bold{b'}}}, \qquad
  s_{\bold{b},\bold{b'}}^{\rm
    CM}=\braket{\psi_{\bold{b}}}{\psi_{\bold{b'}}}\;} with the short
hand notation $\bold{a}\equiv \alpha,l,m$ and $\bold{b}\equiv
\beta,L,M$, the ordinary matrix eigenvalue problem 
\eqn{\bold{H}\bold{\mathbf{C}_i}= \bold{\mathcal{E}_i}
  \bold{\mathbf{C}_i} \;,}
for the configurations remains, where the matrix $\bold{H}$ is given
by
\eqn{H_{k,k'}=\QMa{\Phi_k}{\op{H}}{\Phi_{k'}}\;.}

In order to extend the approach in \cite{cold:gris11} to dipolar
interactions, matrix elements of the type
$\QMa{\varphi_{\bold{a}}}{V_{\rm dd}}{\varphi_{\bold{a'}}}$ have to be 
calculated and added to the relative-motion part of
the Hamiltonian.

\subsection{Basis Set\label{ch:basis}}

The numerical method \cite{cold:gris11} that is extended in this work 
uses spherical harmonics as basis functions for the angular
part of the basis functions. For the radial part of the basis set 
$B$-spline functions $B_{\alpha}(r)$ of order $k$ are used. 
The advantage of using $B$
splines is their compactness in space that leads to sparse overlap and
Hamiltonian matrices. Another relevant property is the continuity of their
derivatives up to order $k-1$.

As a result, the basis functions
\eqn{\phi_{\alpha ,l,m}(\bold{r}) =
  \frac{B_{\alpha}(r)}{r}Y_l^m
  (\theta,\varphi)\;, \label{eq:Basis_rm}}
%
are used to expand the eigenfunctions 
\eqn{\varphi_i=\sum_{\alpha=1}^{N_r}
  \sum_{l=0}^{N_l}\sum_{m=-l}^{l}c_{i; \alpha l m} \; \phi_{\alpha ,l,m} \label{eq:eigenfunc_basis}}
for the relative motion with the expansion coefficients $c_{i, \alpha l m}$. 
The basis sets are characterised by the upper limits $N_l$  of
angular momentum in the spherical-harmonics expansion and the
number $N_r$ of $B$ splines used in the expansion in
\eqref{eq:eigenfunc_basis}. The same type of basis functions 
as in \eqref{eq:Basis_rm} is used for solving for the 
center-of-mass motion functions $\psi$.

The computational effort can be drastically reduced by exploiting
symmetry properties. The Hamiltonian of two atoms interacting via the
interaction potential $V_{\rm int}$ that are trapped in a
$\sin^2$-like or $\cos^2$-like potential oriented along three
orthogonal directions is invariant under the symmetry operations of
the orthorhombic point group $D_{2\rm{h}}$, that are the identity, the
inversion, three two-fold rotations by an angle $\pi$,
and three mirror operations at the Cartesian planes, see
\cite{cold:gris11} for details.

The DDI can be written as
\eqn{V_{\rm dd}(x,y,z)= \frac{C_{\rm dd}}{4\pi} \frac{r^2-3
    z^2}{r^5}\; ,}
with $r=\sqrt{x^2 +y^2+z^2}$ which is also invariant under the
elements of $D_{2\rm{h}}$, since only quadratic orders of $x$, $y$,
and $z$ appear. Therefore, the total Hamiltonian
\eqref{eq:Hamiltonian} remains invariant under the operations in the 
$D_{2\rm{h}}$ symmetry group.

The introduction of symmetry-adapted basis functions allows to treat
each of the eight irreducible representations of $D_{2\rm{h}}$
($A_g$,$B_{1g}$,$B_{2g}$,$B_{3g}$,$A_u$,$B_{1u}$,$B_{1u}$,$B_{1u}$)
independently. This leads to a decomposition of the Hamiltonian matrix
to a sub-block diagonal form which reduces the size of the matrices that 
need to be diagonalized by approximately a factor of 64\footnote{In
  fact, often not all symmetries have to be considered. For example, for
  identical bosons (fermions) only the \textit{gerade} (\textit{ungerade}) 
  ones occur. This leads to a
  further reduction of the numerical efforts.}. For a derivation of the
symmetry-adapted basis functions see \cite{cold:gris11}. They are a
linear combination of non-adapted ones. Hence, for simplicity (but
without loss of generality) we continue the description of the method
using the non-symmetry-adapted basis functions $\phi_i$, while the
numerical implementation uses, of course, the symmetry-adapted ones.

The relative-motion matrix elements that need to be calculated to
extend the existing algorithm toward the DDI are given by
\eqn{\QMa{\phi_{\bold{a}}}{V_{\rm dd}}{\phi_{\bold{a'}}} = \frac{\mu \; C_{\rm dd}}{4 \pi
      \hbar^2}\frac{1}{a_{\rm ho}}
  \QMa{\phi_{\bold{a}}}{\frac{1-3\cos^2(\theta)}{\xi^3}}{\phi_{\bold{a'}}}\;.
\label{eq:V_dd_matrix_elements}}
In \eqref{eq:V_dd_matrix_elements} the dimensionless quantity
$\xi=r/a_{\rm ho}$ with the harmonic-oscillator length $a_{\rm ho} =
\sqrt{\frac{\hbar}{\mu \omega}} $ is introduced.  Furthermore, the
dipole-length $a_{\rm dd} = \frac{\mu C_{\rm dd}}{4 \pi \hbar^2}$ 
characterises the range of the DDI. Expressing the DDI
via the spherical harmonic $Y_2^0$ leads to the matrix elements 
\eqn{\QMa{\phi_{\bold{a}}}{V_{\rm dd}}{\phi_{\bold{a'}}}=
  -\sqrt{\frac{16\pi}{5}} \frac{a_{\rm dd}}{a_{\rm ho}}
    \int_0^{\infty} \frac{B_{\alpha'}(\xi)
    B_{\alpha}(\xi)}{\xi^3}\; d\xi \int_{\Omega} Y_{l'}^{-m'}
  Y_{2}^{0} Y_{l}^{m} d\Omega \;. \label{eq:matrix_element}}
Using the well-known relation
\eqn{\int_{\Omega}Y_{l'}^{m'^{*}}Y_2^0 Y_l^m d\Omega =(-1)^{m'}
  \sqrt{\frac{5(2l'+1)(2l+1)}{4\pi}} \begin{pmatrix} l^{\prime} & 2 &
    l \\ 0 & 0 & 0 \end{pmatrix}
  \begin{pmatrix} l^{\prime} & 2 & l \\ -m^{\prime} & 0 &
    m \end{pmatrix}\label{eq:DD_coupling_elements}}
between spherical harmonics and the Wigner-$3J$ symbols, it is evident
that the dipole-dipole coupling elements in
\eqref{eq:DD_coupling_elements} vanish, except if the following three
conditions are fulfilled simultaneously:\\
(i) the sum of the $l$ quantum numbers is even, i.\,e., $l' +l +2 =
2n$ with $n \in \mathbb{N}$,\\
(ii) $|l' -l|\leq 2 \leq l' + l$, which refers to the triangular 
inequality, and\\
(iii) the sum of the $m$ quantum numbers needs to be zero, i.\,e.,
$-m'+m=0$.\\
From the third condition it follows that $m$ remains a good quantum
number, i.\,e. $[\op{L}_z,\op{H}] =0$. The product of Wigner-$3J$
symbols in \eqref{eq:DD_coupling_elements} can be calculated in an
extremely accurate and efficient way.
 
Clearly, the DDI adds additional numerical demands to the problem,
since already at the level of solving the Schr\"odinger equation of
the relative-motion Hamiltonian $\op{h}_{\rm rm}$ a coupling of all
even or all odd $l$ quantum numbers is introduced. This increases the
number of non-zero matrix elements in comparison to the case without
DDI significantly.

\section{Results}
\label{ch:results}
In this section we present results of the solution of the Schr\"odinger
equation with the Hamiltonian of \eqref{eq:Ham_rm} for different
trapping potentials.

For the specific trapping potential, the mass of $^7\rm{Li}$ is used for 
the masses of the dipolar particles $m_1$ and $m_2$. Additionally, the 
polarizability of the dipolar particles $\alpha= 200 \; a. u.$ is chosen. 
Furthermore, the laser parameters, the wave length $\lambda=1000 \; \rm{nm}$ 
and the intensity $I=1000 \; \rm{W}/ \rm{cm}^2$, which characterise the trapping 
potential are used. The resulting trapping frequency is $\omega= 152,2 \; \rm{KHz}$ 
(for $x$,$y$ and, $z$ direction in the case of an isotropic trap ).

\begin{figure}[htpb]
  \includegraphics[width=\linewidth]{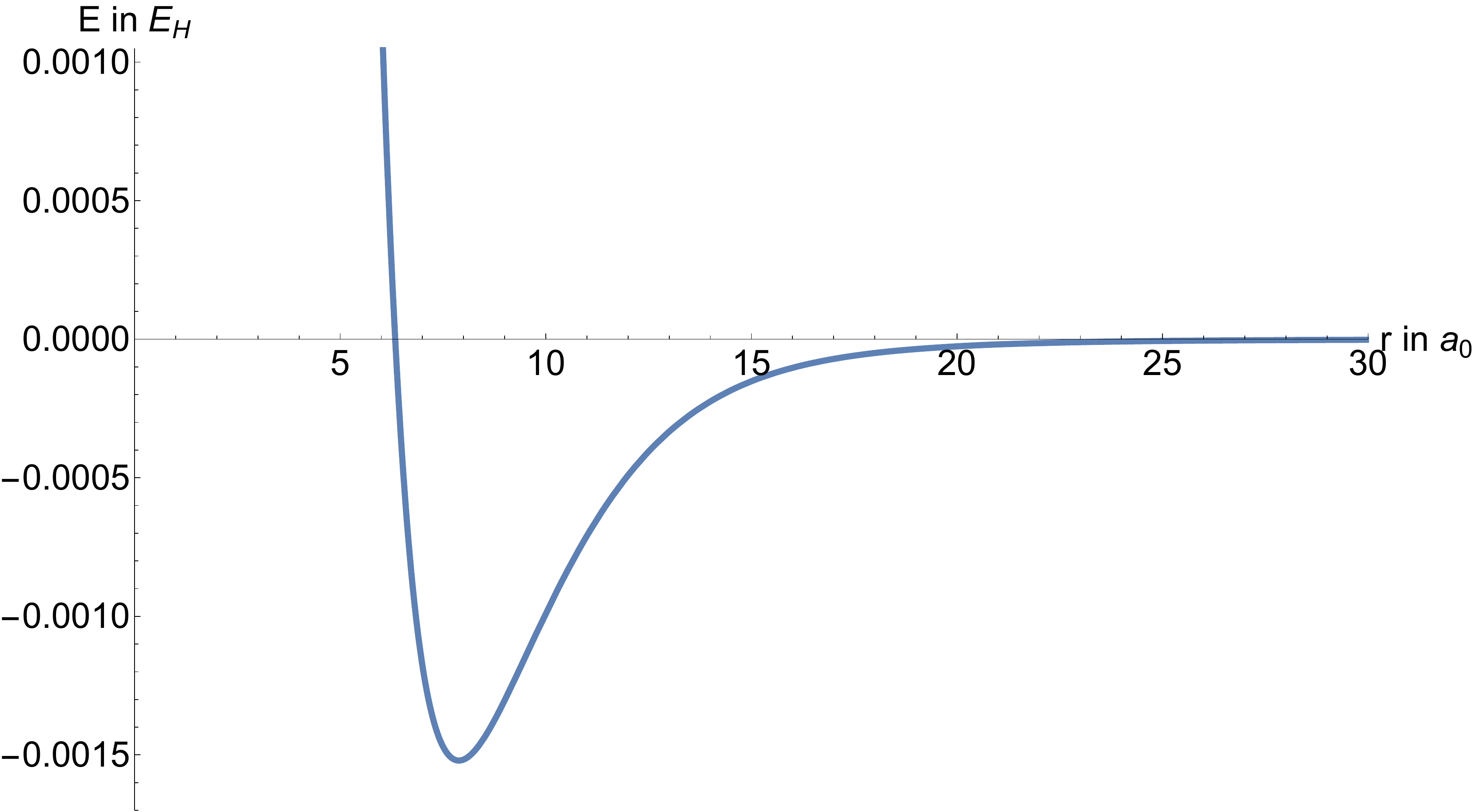}
  \caption{Interaction potential of two $\text{\rm Li}$ atoms in the
    lowest triplet state. This atomic interaction potential is used as a
    prototype realistic atom-atom interaction potential.}
  \label{Fig:Li_pot}
\end{figure}

As a {\it generic} example for a realistic short-range interaction potential, in the
present study, the one of two $\rm{Li}$ atoms in their lowest triplet 
state $a\, ^3\Sigma_{\rm u}^+$ \cite{cold:gris07} is chosen which is shown in
\figref{Fig:Li_pot}. This interaction potential of lithium is
numerically not too demanding since it provides a smaller number of bound
states than, e.\,g., the one of Cs, Cr, Dy, or Er, and hence, 
a smaller number of $B$ splines yields converged results. 

In the ultracold regime, the isotropic short-range interaction can be
parameterised by the $s$-wave scattering length $a_{\rm sc}$ which is
determined by the energy of the most weakly bound state. In order
to simulate, e.\,g., the variation of the scattering length in the
vicinity of a magnetic Feshbach resonance or a different system 
of particles, the approach described in
\cite{cold:gris10} is used where a small modification of the inner wall of the
potential varies the position of the last bound state and hence
the $s$-wave scattering length in the absence of the DDI. 

However, it is important to note that the concept of the $s$-wave scattering
length breaks down for a non-zero dipole moment. First, a partial-wave
expansion does not decouple the wavefunction with respect to the
angular momentum quantum number $l$, since the DDI couples all even
(odd) $l$ quantum numbers. Second, the $1/r^3$ tail of the DDI leads
besides the usual linear term also to a logarithmic term in the
asymptotic part of the wavefunction \cite{dipol:sinh07}. This
logarithmic behaviour cannot be described by short-range $s$-wave scattering.

\subsection{Isotropic harmonic trapping potential\label{ch:harmonic}}

First, we consider an isotropic harmonic confinement. The harmonic
potential
\eqn{V_i(\bold{r_i})=\sum_{j=x_i,y_i,z_i} V_{0,j}k_j^2 j^2 \;}
is obtained by a Taylor expansion of the optical-lattice potential
truncated at second order. Introducing the harmonic oscillator
frequencies $\omega=\sqrt{\frac{2V_0 k^2}{\mu}}$ and
$\Omega=\sqrt{\frac{4V_0 k^2}{M}}$ where $M=m_1+m_2 = 2m$, the
potential
\eqn{V_i(\bold{r_i})&=V_{\rm rm}(r) \; \; + \;V_{\rm CM}(R) =
  \frac{1}{2}\mu \omega^2 r^2 + \frac{1}{2} M \Omega^2 R^2}
is separable in relative and center-of-mass coordinates, i.\,e. the
coupling term $\op{W}(\bold{r},\bold{R})$ vanishes. Since additionally
the DDI affects only the relative-motion coordinates, the
center-of-mass Hamiltonian is the one of an ordinary harmonic
oscillator. Thus, we concentrate on the relative-motion Hamiltonian in
\eqref{eq:Hamiltonian_rm}. 

The total energy spectrum can be characterised by two different energy
regimes. In the \textit{bound-state regime}, i.\,e.\ the energy range 
below the dissociation threshold in the absence of a trap, the 
characteristic energies are
on the order of the energies of the interaction potential
$V_{\mathrm{sh}}$ that supports bound rovibrational states. In the
\textit{trap-state regime}, i.\,e.\ for the states above the dissociation 
threshold in the absence of a trap, the characteristic energies are on the order of
the trap-discretized continuum states that we denote as trap
states in the following. In our case, the typical trap-state energies are of the order of a 
few $\hbar \omega$ which corresponds in atomic units to about
$10^{-12}E_H$.  The characteristic depth of the short-range potential 
$V_{\rm sh}$ is about $-10^9 \hbar
\omega$. The typical energy difference of the vibrational levels in
units of $\hbar \omega$ is approximately $10^8 \hbar \omega$. Also the
characteristic energy difference of the rotational energy levels of
about $10^7 \hbar \omega$ is orders of magnitude larger compared with
the characteristic energy scales of a few $\hbar \omega$ in the 
trap-state regime. Therefore, the two regimes will be discussed 
separately in the following two subsections.

\subsubsection{Bound-state regime.} 
\label{sec:bound_regime}

First the bound-state regime is considered. Since we adopt a realistic
interaction potential there exist more than one bound state. These bound 
states can
couple to each other due to the DDI.

\begin{figure}[htpb]
  \includegraphics[width=\linewidth]{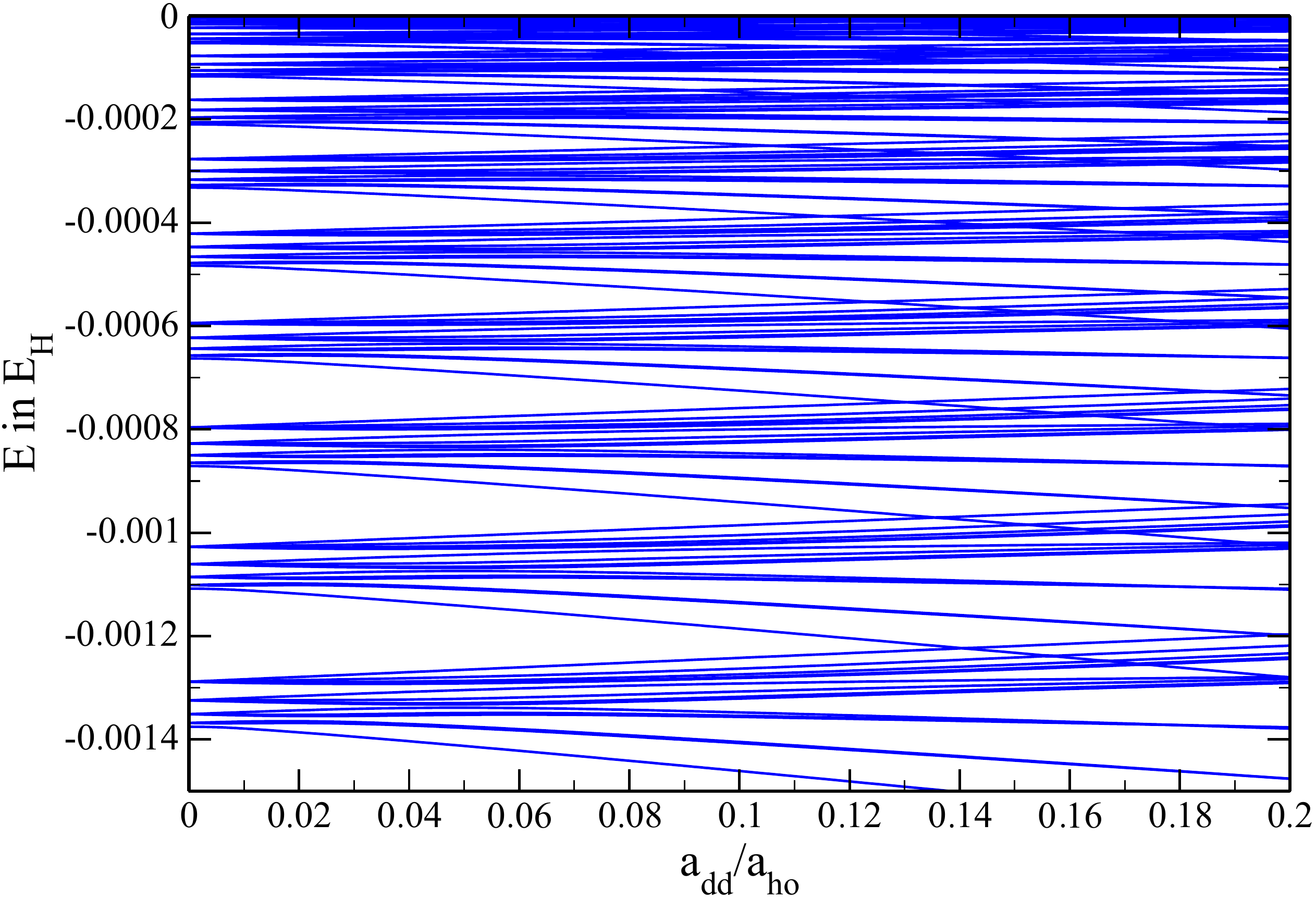}
  \caption{Relative-motion energy spectrum for the $A_g$ symmetry of the
    $D_{2h}$ point group showing the bound regime of the energy
    spectrum for variable dipolar interaction strength.}
  \label{Fig:energy_spec_bound}
\end{figure}

\figref{Fig:energy_spec_bound} shows the energy spectrum of the $A_g$
symmetry of two identical dipolar particles in an isotropic harmonic
trap interacting via the short-range interaction potential 
(\figref{Fig:Li_pot}) as a function of the dipole interaction strength of the
DDI, which is characterised by the ratio between the dipole-length
$a_{\rm dd} = \frac{\mu C_{\rm dd}}{4 \pi \hbar^2}$ and the harmonic
oscillator length $a_{\rm ho}$.  Since the $A_g$ symmetry is
\textit{gerade}, the spectrum represents identical bosons. The dipolar
interaction strength $\frac{a_{\rm dd}}{a_{\rm ho}}$ determines the
behaviour of the system in the long-range regime. 
In \figref{Fig:energy_spec_bound} groups of states appear which are
partly degenerate at $\frac{a_{\rm dd}}{a_{\rm ho}}=0$ and begin to
separate for increasing dipole interaction strength.
Each group of states corresponds to one vibrational energy level and
its rotational excitations. In total there are eleven vibrational
states supported by the short-range potential shown in
\figref{Fig:Li_pot}. In the calculation the number of rotational excitations for each
vibrational energy level is limited by the number $N_l$ of the basis
set in \eqref{eq:Basis_rm}. 

The group of states in the energy interval between $E=-0.0014 \; E_H$
and $E=-0.0012 \; E_H$ corresponds to the vibrational ground state and
its rotational excitations.  The next set of states between $E=-0.0012
\;E_H$ and $E=-0.0010 \;E_H$ corresponds to the first excited vibrational
state and its rotational energy levels. For each set of
rovibrational states the properties are similar. As is visible from
\figref{Fig:energy_spec_bound}, the different rovibrational states of
each set respond differently to the increasing dipole interaction strength.
Since the head-to-tail configuration corresponds to states with the
rotational quantum numbers $m=0$, these states decrease in energy with
increasing dipole interaction strength. The states in the
side-by-side configuration correspond to $l = |m| \neq 0$ quantum
numbers and increase in energy for increasing dipole interaction strength.
The rovibrational ground state at $E \approx -0.00137\; E_H$ changes
most strongly with the DDI, since the expectation value of
$\bar{r}=\QMa{\varphi}{r}{\varphi}$ is the smallest and therefore the
DDI matrix elements are the largest, because the dipole-dipole matrix
elements scale as $r^{-3}$. With increasing vibrational
excitation $\bar{r}$ gets larger and the splitting of the different
rovibrational states is weaker.  
\begin{figure}[htpb]
  \includegraphics[width=\linewidth]{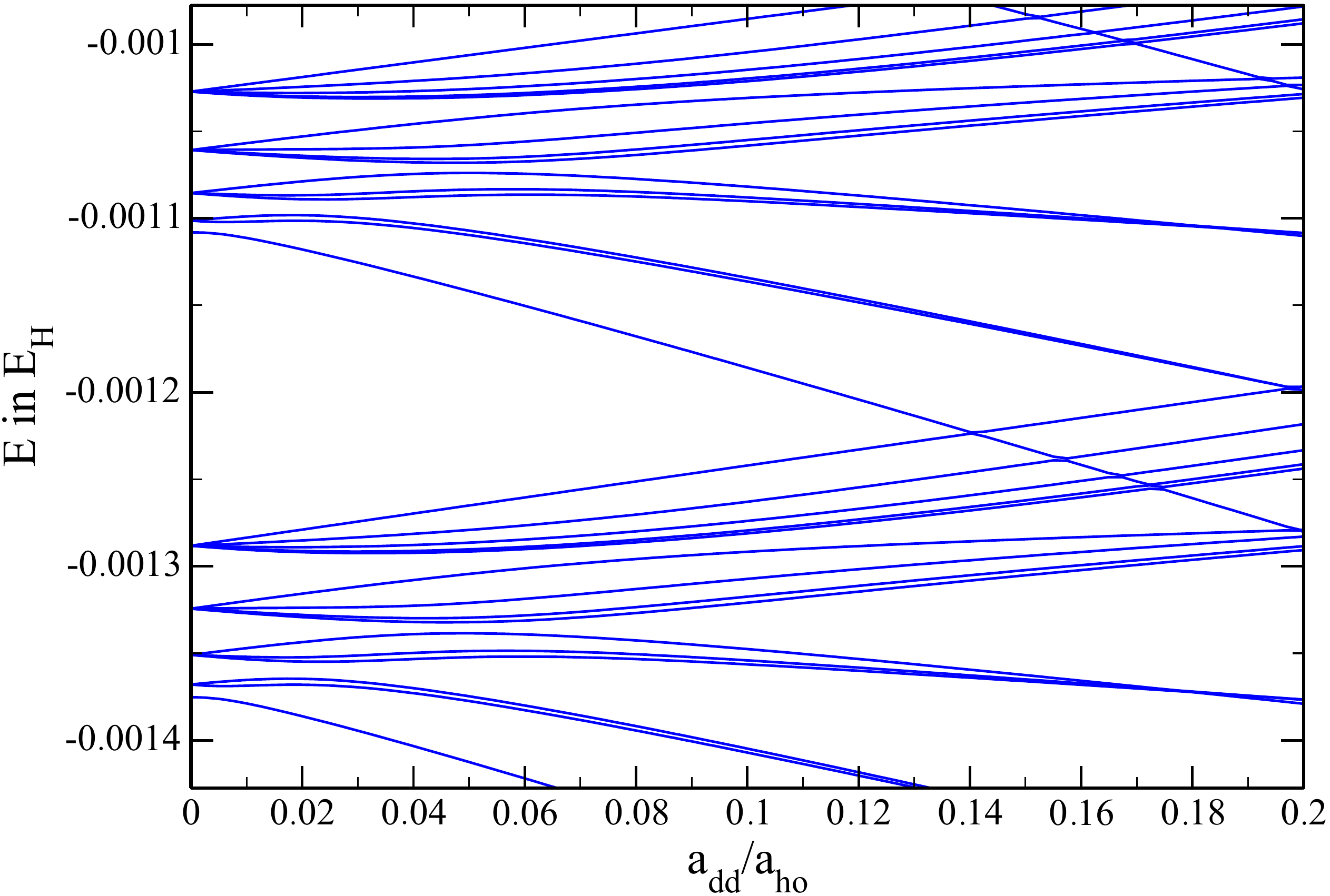}
  \caption{Magnified view on \figref{Fig:energy_spec_bound} that shows the
    different behaviour of the rovibrational levels with increasing
    dipole interaction strength.}
  \label{Fig:energy_spec_bound_zoom}
\end{figure}

\begin{figure}[ht]
\hspace*{-7mm}
\includegraphics[width=1.1\linewidth]{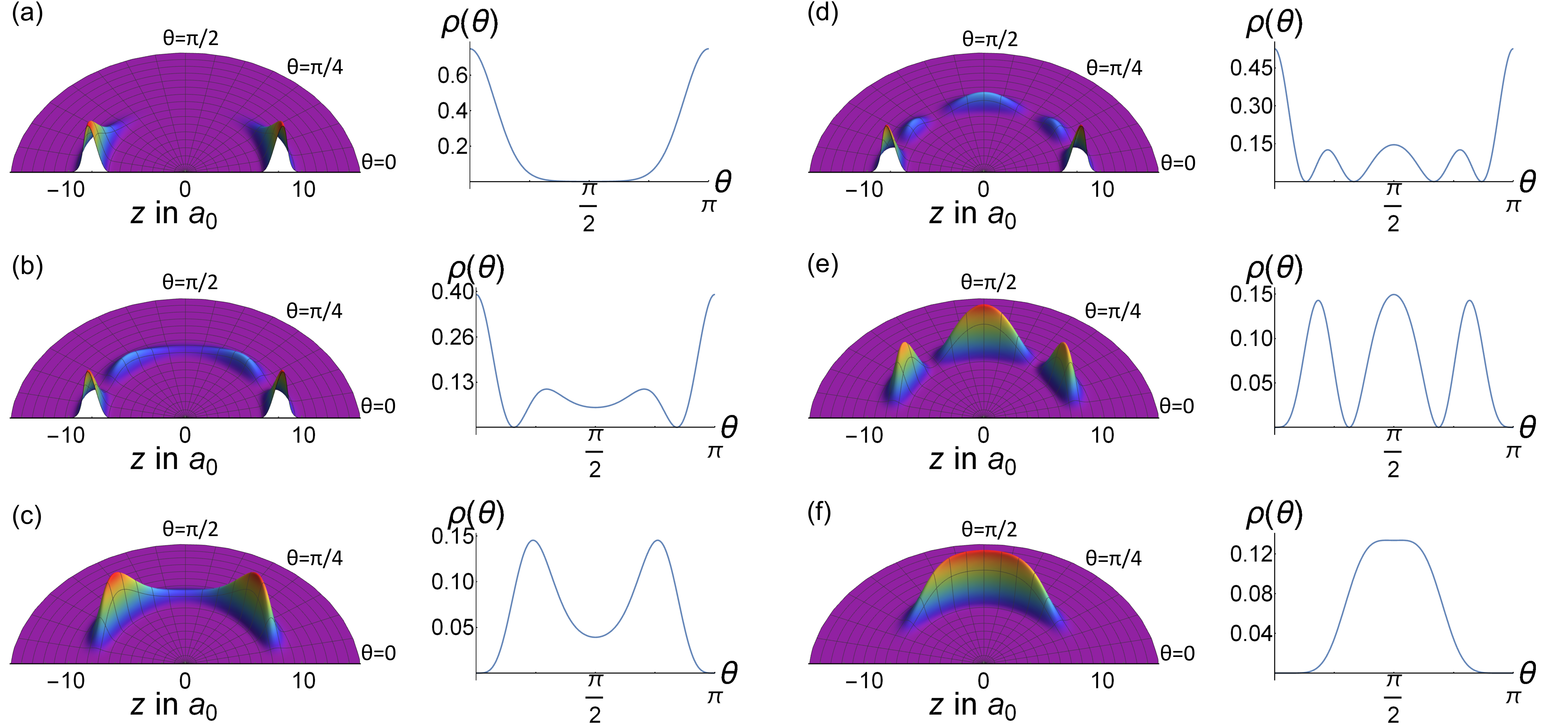}
  \caption{Pair densities $\rho(r,\theta,\phi = 0)$ of the 6
    energetically lowest lying bound states (sorted from (a) to (f)) for the
    dipole interaction strength $\frac{a_{\rm dd}}{a_{\rm ho}}=0.025 $. The
    solutions of the full Hamiltonian in \eqref{eq:Ham_rm}
    (3D plots) are compared to the ones of the model Hamiltonian
    in \eqref{eq:mod_Hamiltonian} (2D plots).  }
  \label{Fig:bound_comp}
\end{figure}

In \figref{Fig:energy_spec_bound_zoom} the different splittings of the
first two sets of rovibrational states are shown in an enlarged view
compared to \figref{Fig:energy_spec_bound}. In the present calculation 
the basis set is includes only l quantum numbers up to $N_l=8$. Since 
only even numbers of $l$ and $m$ are allowed in the
\textit{gerade} $A_g$ case, there are five values of $l$ contained in the 
basis and the number of rotational states per vibrational level visible 
in \figref{Fig:energy_spec_bound_zoom} is correspondingly limited 
to this number. The $m$
degeneracy of the rotational levels is lifted because the DDI breaks
the spherical symmetry. From an analysis of the wavefunction of the
deeply bound states the general properties of the states becomes
evident, see \figref{Fig:bound_comp}. While the $m$ quantum number is
preserved, the $l$ quantum numbers are coupled by the DDI. Therefore,
each state consists of a fixed $m$ quantum number with contributions
from all even $l$ quantum numbers. On this basis it is possible to
judge whether a state belongs dominantly to the head-to-tail or the
side-by-side configuration or is in between these extremes. In general, the
states with $m=0$ correspond to the classical head-to-tail
configuration and states with $|m|>0$ represent more the side-by-side
configuration, especially the states with $l=|m| \neq 0$ have a
pronounced side-by-side geometry.

In \figref{Fig:bound_comp} the pair densities of the vibrational
ground state with its rotational excitations are shown. These pair
densities do not possess a vibrational excitation as can be seen from
the missing nodes in the radial part of the pair densities. For the
excited rotational states \figref{Fig:bound_comp}(b) to (f) additional
nodes appear in the angular part of the pair densities.  Moreover, the
states in \figref{Fig:bound_comp}(a), (b), and (d) corresponds to the
head-to-tail configuration, since the probability for two dipolar
particles to stand on top of each other is the largest.  In contrast, the
state in \figref{Fig:bound_comp}(c) does not show a clear side-by-side 
or head-to-tail configuration as it has minima in both cases. The state  
in \figref{Fig:bound_comp}(f) has a maximum  for the 
side-by-side configuration and increases in energy for an increasing
dipole interaction strength, as described above and visible from
\figref{Fig:energy_spec_bound_zoom}. Finally, the state shown in 
\figref{Fig:bound_comp}(e) has a node for the head-to-tail configuration 
and some maximum for the side-by-side configuration, but equal maxima 
for the angle in between.

\begin{figure}[htpb]
  \centering \includegraphics[width=\linewidth]{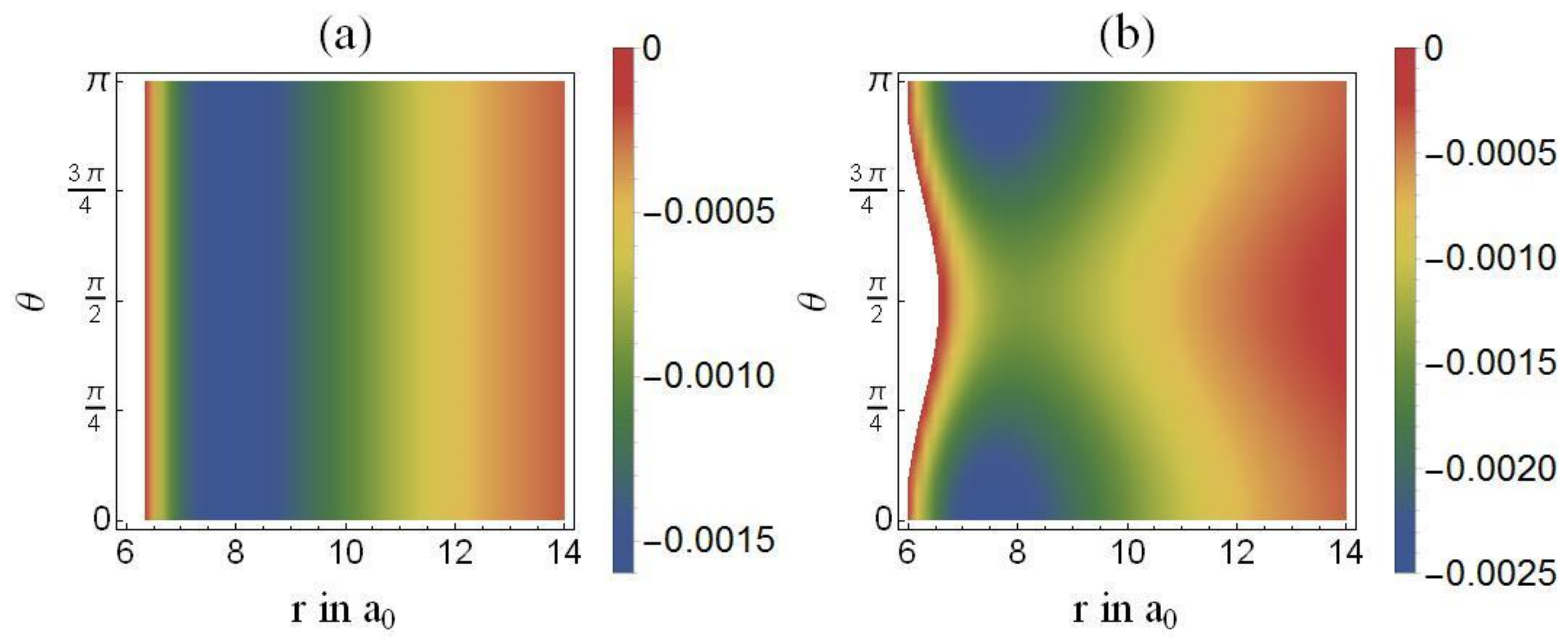}
  \caption{Full interaction potential $V_{\rm int}$ for different interaction
    strengths $\frac{a_{\rm dd}}{a_{\rm ho}}$. (a) Isotropic case
    without DDI ($\frac{a_{\rm dd}}{a_{\rm ho}}=0$).(b) Short-range
    potential with a non-zero DDI ($ \frac{a_{\rm dd}}{a_{\rm
        ho}}=0.15$).}
  \label{fig:potential_cont}
\end{figure}

\figref{fig:potential_cont} shows how the total potential 
$V_{\rm int}$ changes for small particle separations due to the 
dipolar interaction ($\frac{a_{\rm dd}}{a_{\rm ho}}=0$ or 
$\frac{a_{\rm dd}}{a_{\rm ho}}\neq 0$ ). The total potential in
\figref{fig:potential_cont}(b) becomes increasingly anisotropic due to
the DDI and dips for the head-to-tail configuration as well as bumps 
for the side-by-side configuration appear. Since the total potential in
\figref{fig:potential_cont}(b) is shallower in the $\theta$ direction
than in the radial direction, it is more likely to have first
excitations in the angular part of the pair densities, see
\figref{Fig:bound_comp}.

In order to better understand the behavior of the pair densities shown
in \figref{Fig:bound_comp} we consider the model Hamiltonian
\eqn{\op{H}_{\rm model}=\frac{\op{L}^2}{2\mu r_0^2}+ \frac{C_{\rm
      dd}}{4\pi}
  \frac{1-3\cos^2(\theta)}{r_0^3}\label{eq:mod_Hamiltonian}}
consisting only of the angular part of the full Hamiltonian
\eqref{eq:Hamiltonian} for constant $r=r_0$. $r_0$ is chosen such that
the pair density of the vibrational ground state has its maximum. The
solution of the model Hamiltonian in \eqref{eq:mod_Hamiltonian} is
obtained by diagonalizing the model Hamiltonian in the basis of
spherical harmonics.

\begin{figure}[htpb]
  \includegraphics[width=\linewidth]{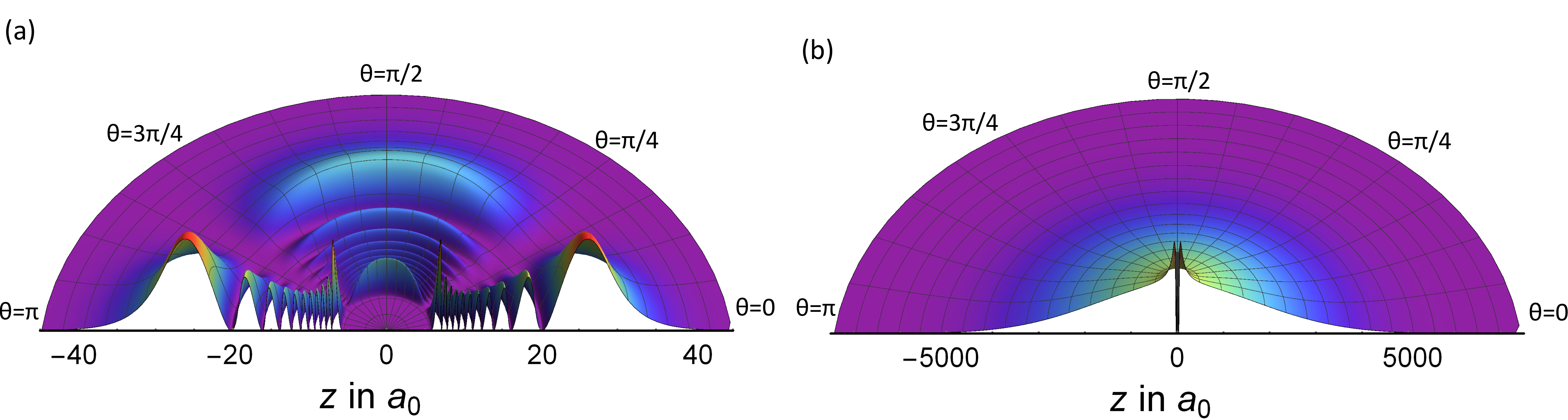}
  \caption{Relative-motion pair density $\rho (r,\theta ,0)$ of 
    (a) a bound state close to the interaction potential threshold with $E =
    -49,283 \; \hbar \omega$ at $\frac{a_{\rm dd}}{a_{\rm ho}}= 0.01$ and (b)
    the first trap state with $E = 1.488 \; \hbar \omega$ for a dipole
    interaction strength of $\frac{a_{\rm dd}}{a_{\rm ho}}= 0.001$}
  \label{Fig:boundstate147}
\end{figure}

In \figref{Fig:bound_comp}, very good agreement is visible comparing
the full solution (3D plots) for constant $r_0 \approx 8$ with the
solution of the model Hamiltonian (2D plots). However, higher excited
vibrational states possess a more pronounced $r$ dependence which leads to a
more complex radial nodal structure as shown in
\figref{Fig:boundstate147}(a). In this case the simplified model Hamiltonian is
hence not expected to reproduce well the radial behaviour.

Additionally, there exist \textit{metastable states} above the
trap-free dissociation threshold. In the absence of a DDI such states 
gain their stability by the centrifugal barrier, but they can 
dissociate by tunnelling through this barrier. With increasing dipole 
interaction strength those metastable states
strongly respond to the DDI, since the distance of the two dipolar
particles is small compared to the inter-particle distance of trap states,
see \figref{Fig:boundstate147}. For increasing DDI, a metastable state
can increase (decrease) in energy depending on whether the
configuration of the state is dominantly side-by-side (head-to-tail). 
Similarly, a bound state that is in energy close to the
trap-free dissociation threshold for a vanishing DDI can increase in
energy above the threshold if it has a predominant side-by-side
configuration. In this case the state can surpass the (trap-free) 
continuum threshold and becomes an unbound or metastable state.

\subsubsection{Trap-state regime.}
Next, we consider the trap-state regime. %
\begin{figure}[htpb]
  \includegraphics[width=\linewidth]{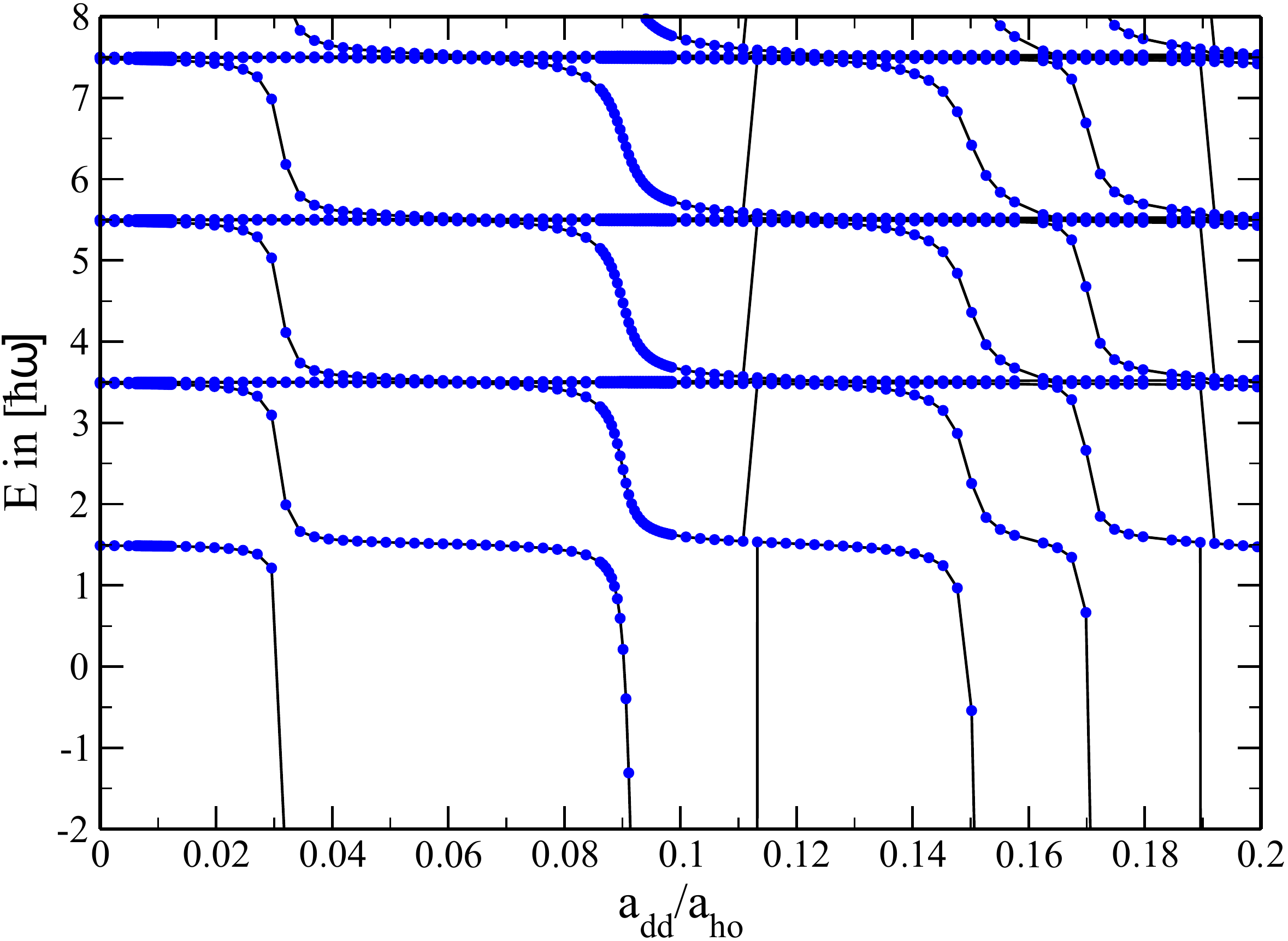}
  \caption{Relative-motion energy spectrum for the $A_g$ symmetry of the
    $D_{2h}$ point group showing the trap-state regime of the energy
    spectrum as a function of the dipole interaction strength. }
  \label{Fig:energy_spec}
\end{figure}
The basically horizontal lines in \figref{Fig:energy_spec} correspond to the
harmonic-oscillator trap states in the diabatic picture. Since this
is the energy spectrum of the $A_g$ symmetry, only even numbers of $l$
and $m$ are allowed. The energy of two non-interacting particles in a
three-dimensional isotropic trap in spherical coordinates is $E_{\rm
  ho}= (k + 3/2) \hbar \omega$, with $k\equiv 2n+l$. Thus the
degeneracy of the energy level $k$ is $(k+2)(k+4)/8$ where $k=0,2,4,6...$ are
only even numbers due to the $A_g$ symmetry. For the even $k$ quantum
numbers the spacing of the harmonic-trap states is $2 \hbar \omega$
in accordance with the energies shown in \figref{Fig:energy_spec}.

The almost vertical lines at $\frac{a_{\rm dd}}{a_{\rm
    ho}}=0.03$, $\frac{a_{\rm dd}}{a_{\rm ho}}=0.09$, $\frac{a_{\rm
    dd}}{a_{\rm ho}}=0.15$, and $\frac{a_{\rm dd}}{a_{\rm ho}}=0.17$
are bound states that show avoided crossings with the 
trap-discretized continuum states leading to dipole-induced resonances, see
\secref{sec:dir}. These bound states consist of a mixture all
spherical harmonics with different $l$ but equal $m$ quantum numbers.
The fact that those bound states appear as almost vertical lines 
is a consequence of the very different energy scales differing by 7 to 
8 orders of magnitude between the bound states 
and the trap states, as was discussed earlier.

The bound state at $\frac{a_{\rm dd}}{a_{\rm ho}}=0.11$ lies rather 
close to the dissociation threshold for vanishing DDI and has 
a predominant side-by-side configuration. This leads to an increasing 
energy that surpasses the (trap-free) dissociation threshold for 
a large DDI. The bound state at $\frac{a_{\rm dd}}{a_{\rm ho}}=0.19$ is 
a metastable state, see \secref{sec:bound_regime}.  Both states 
do not couple to trap states because of non-agreeing $m$ quantum numbers. 

\FloatBarrier

\subsubsection{Dipole-induced resonances.}
\label{sec:dir}
A key feature of the energy spectrum in \figref{Fig:energy_spec} is 
the occurrence of broad 
scattering resonances, which we denote as dipole-induced resonances
(DIR). These resonances manifest as broad avoided crossings in the
energy spectrum and are found for 
$\frac{a_{\rm dd}}{a_{\rm ho}} \approx 0.03$, $\frac{a_{\rm
    dd}}{a_{\rm ho}} \approx 0.09$, $\frac{a_{\rm dd}}{a_{\rm ho}}
\approx 0.15$, and $\frac{a_{\rm dd}}{a_{\rm ho}}\approx 0.17$. They can be
understood in the following way. Increasing the dipole interaction
strength deepens the interaction potential $V_{\rm int}$ and 
introduces a new bound state in the inter-particle potential. This bound
state crosses in energy with trap states. Due to the dipolar coupling
these crossings become avoided.  A further increase of the interaction
strength repeats this process leading to the series of resonances
(avoided crossings) visible in \figref{Fig:energy_spec}. The
resonances shown in \cite{dipol:kanji08, dipol:shi12, dipol:rone06,dipol:bort06}
seem also to be DIR.

The excited trap states that do not couple to the bound states
(e.\,g. the almost horizontal line at $E \approx 3.5 \hbar \omega$)
are $l \neq 0$ states. The centrifugal barrier $\frac{\hbar^2}{2\mu}\frac{l(l+1)}{r^2}$ 
shields the trap states from noticing the bound state that is much more confined 
compared to a trap state, see \figref{Fig:boundstate147}. Due to the 
shielding effect of the centrifugal barrier for the trap states with 
$l \neq 0$ there is very little overlap with the bound state. This  
leads to an almost vanishing coupling.

\FloatBarrier

\subsubsection{Dipolar coupling of trap states.}

\begin{figure}[htpb]
  \centering \includegraphics[width=\linewidth]{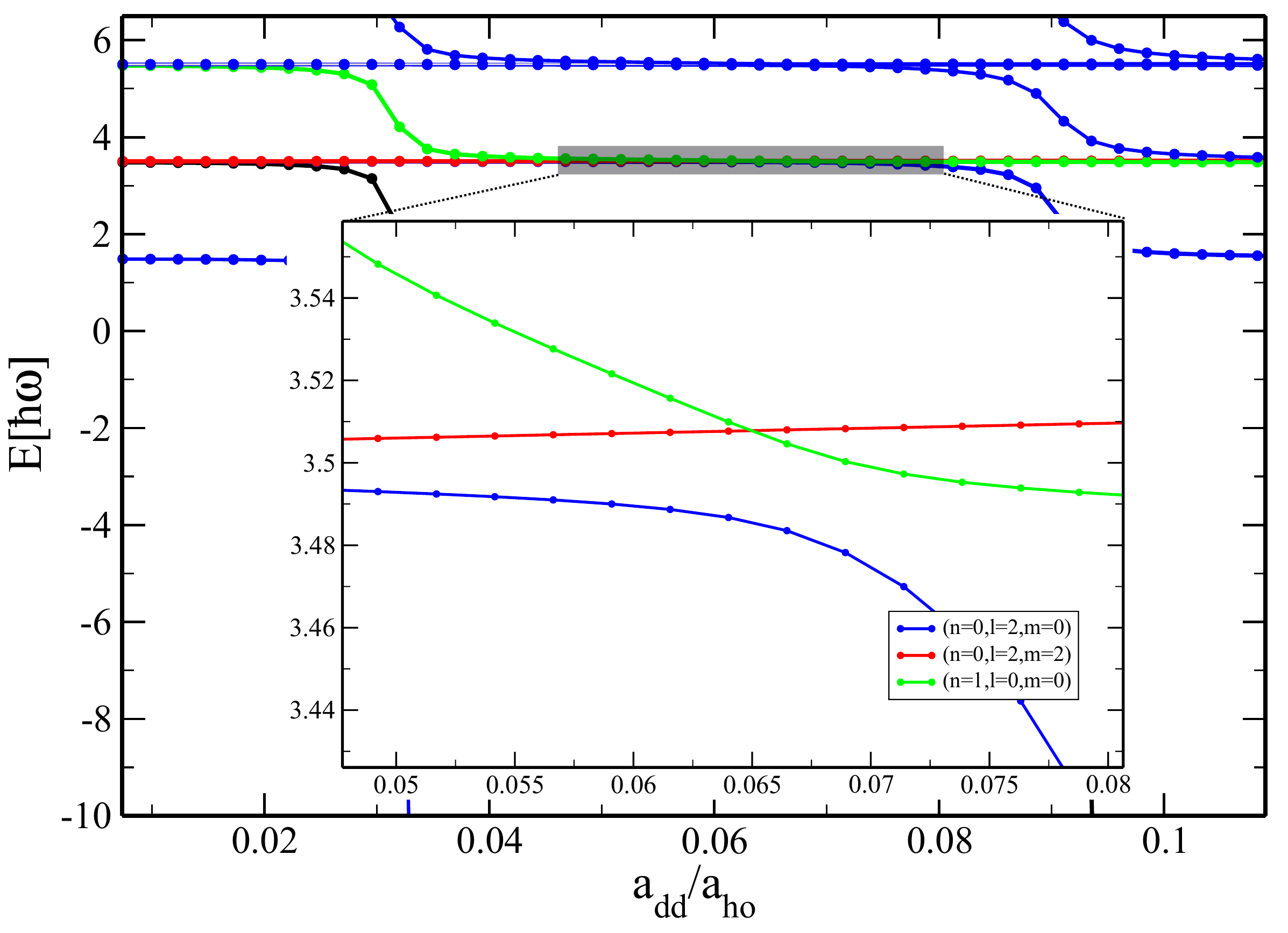}
  \caption{Magnified view on \figref{Fig:energy_spec} that shows an
    avoided crossing between trap states. The quantum numbers in the
    legend are associated with the quantum numbers of a harmonic
    oscillator. The indexing of the states with the quantum numbers  
    is only valid far away from the resonance.  }
  \label{fig:energy_zoom}
\end{figure}

An enlarged view of \figref{Fig:energy_spec} is shown in
\figref{fig:energy_zoom}. The red state corresponds to a trap state
where the dipoles possess a side-by-side configuration. It is well
described by a harmonic oscillator state with
$(n=0,l=2,m=2)$. Therefore, the energy of this state increases with
the dipole interaction strength. Moreover, there is a non-avoided crossing of
the red and the green state. The green state is a trap state with
quantum numbers $(n=1,l=0,m=0)$. From the selection rules of the DDI
(see \secref{ch:basis}) it follows that only states with equal $m$
quantum numbers can couple due to the DDI. Consequently, this is a 
non-avoided crossing of two non-interacting states. Furthermore, there
is an avoided crossing between the blue state $(n=0,l=2,m=0)$ and the
green state. This is a direct consequence of the selection rules of the
DDI. The coupling of the trap states is significantly smaller than the
coupling in the DIR, because the coupling matrix element
\eqref{eq:matrix_element} for two trap states is smaller than for a trap
and a bound state. Hence, the coupling matrix element is larger 
for small $r$, which is the case for the much more strongly confined 
bound state compared with a trap state. The avoided crossings between 
trap states can nevertheless be used to 
adiabatically transfer an isotropic trap state to a polarised one 
and could thus be used for some control scheme. 

\FloatBarrier

\subsubsection{Influence of the short-range potential.}

\begin{figure}[htbp]
  \includegraphics[width=\linewidth]{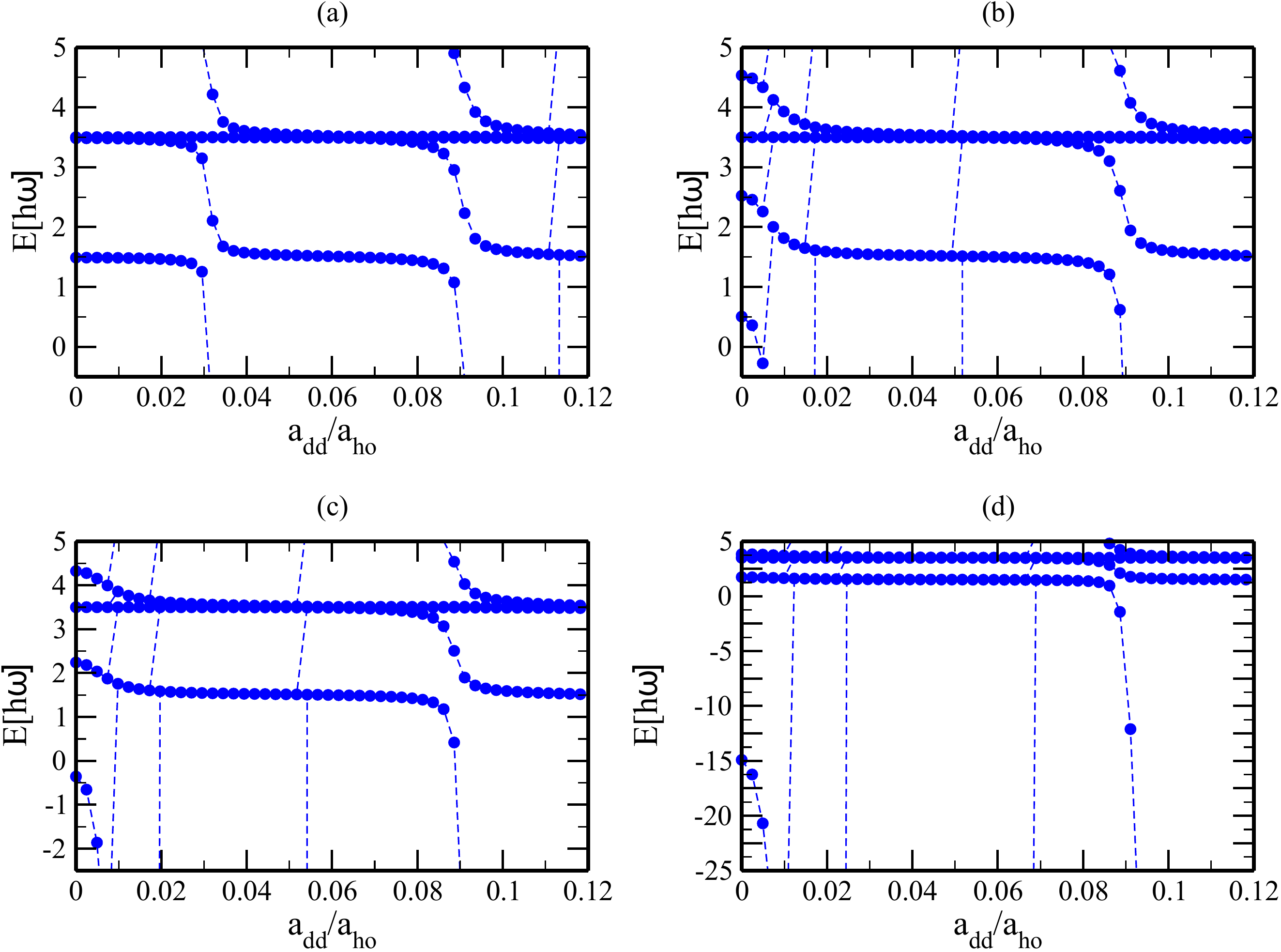}
  \caption{Relative-motion energy spectra for different short-range
    potentials which correspond to different $s$-wave scattering length
    $a_{\rm sc}$ for zero dipole interaction strength. This corresponds 
    to either an attractive ($a_{\rm sc}<0$) or a repulsive ($a_{\rm sc}>0$) 
    short-range interaction. Energy spectrum for (a) $\frac{a_{\rm ho}}{a_{\rm
        sc}} = -100$,  (b) $\frac{a_{\rm
        ho}}{a_{\rm sc}} = 0.01$,  (c) $\frac{a_{\rm ho}}{a_{\rm sc}} = 1.0$,  
        and (d) $\frac{a_{\rm ho}}{a_{\rm sc}} = 5.0$.}
  \label{fig:d_y_over_a}
\end{figure}

To investigate the influence of the short-range potential for dipolar 
particles,
the $s$-wave scattering length of the short-range potential for zero
DDI is varied by an inner-wall manipulation \cite{cold:gris10}. 
The energy spectra of two dipolar particles interacting via different
short-range potentials are shown as a function of the dipole interaction strength 
in \figref{fig:d_y_over_a}. Each energy spectrum refers to a
different $s$-wave scattering length of the bare short-range potential
$V_{\mathrm{sh}}$.

As in \figref{Fig:energy_spec} the horizontal lines correspond to
harmonic-trap states.  The variation of $V_{\mathrm{sh}}$ changes the
position of the DIR as is shown in \tabref{tab:Position}. %
\begin{table}
\caption{\label{tab:Position}Position of the DIRs in
  \figref{fig:d_y_over_a}.}
\begin{indented}
\item[]\begin{tabular}{@{}llll} \br Scattering length $a_{\rm sc}\,
    [a_0]$& Position of the &Position of the \\ of $V_{\mathrm{sh}}$ &
    first DIR
    [$a_{\rm dd}/a_{\rm ho}$]& second DIR [$a_{\rm dd}/a_{\rm ho}$] \\
    \mr -25 & 0.03 & 0.09\\ 250000 &0.005&0.09\\ 25000 &0.005&0.09\\
    500 & 0.004&0.09\\ \br
\end{tabular}
\end{indented}
\end{table}
The short-range potential has a strong influence on the position of
the first DIR because for small dipole interaction strength $\frac{a_{\rm
    dd}}{a_{\rm ho}}$ the DDI is negligible. In contrast, 
the position of the second DIR (at $\frac{a_{\rm dd}}{a_{\rm ho}} = 0.09$) is
almost unaffected by a manipulation of $V_{\mathrm{sh}}$.
For increasing dipole interaction strengths
$\frac{a_{\rm dd}}{a_{\rm ho}}$ the DDI becomes dominant compared to
the short-range potential $V_{\mathrm{sh}}$. Consequently, the
long-range behavior of the wavefunction is not longer affected by changes in
$V_{\mathrm{sh}}$. This can be
understood on the basis of the different energy shifts for bound states
resulting from an inner-wall manipulation compared to the one induced
by increasing the DDI. While the energy of the bound states respond 
sensitively on an inner-wall variation for small DDI, for a large DDI
the energy shift due to the inner-wall manipulation is negligible
compared to the changes in the vibrational energy levels induced by a
variation of the DDI, see, e.\,g., \figref{Fig:energy_spec_bound}. 
It is certainly remarkable that this effect can be used to directly change the
position of the first DIR, which allows for controlling and
manipulating dipolar quantum gases.

\subsection{Anisotropic sextic trapping potential\label{ch:sextic}}

Every realistic potential is finite and hence certainly
anharmonic. Therefore, in the following, the harmonic approximation is
abandoned and a sextic potential is considered which has proven to
accurately describe center-of-mass to relative motion coupling in
single-well potentials \cite{cold:gris09,cold:sala12,cold:sala13}. A  
pancake-shaped trap with an anisotropy of
$\frac{\omega_z}{\omega_{\bot}} = 10$ is adopted. Such a quasi-2D
geometry is a common experimental set-up to stabilise a dipolar BEC
against collapse \cite{dipol:koch08}. Again, the short-range potential
of Li is chosen as a prototype short-range potential $V_{\mathrm{sh}}$.
The sextic potential 
\eqn{V_i(\bold{r_i}) = \sum_{j=x_i,y_i,z_i} \, V_j \left( k_j^2 j^2 -
  \frac{1}{3} \, k_j^4 j^4 + \frac{2}{45} \, k_j^6 j^6 \right)\;}
is obtained by a Taylor expansion of a $\sin^2$ optical lattice
up to the sixth order.
While in the harmonic case center-of-mass and relative motion decouple, 
for a sextic potential the coupling term 
\eqn{\op{W}(\bold{r},\bold{R})=\sum_{c=x,y,z} V_{0,c}\left(-k_c^4
  r_c^2 R_c^2 + \frac{1}{3}k_c^6 r_c^2 R_c^4 + \frac{1}{12} k_c^6
  r_c^4 R_c^2\right)\; \label{eq:sextic_coupling}}
does not vanish. Note, the coupling term is non-zero even for a 
harmonic confinement if the two particles have different products of
polarizability and mass. Hence, for the description of the full
spectrum the Schr\"odinger equation of the six-dimensional Hamiltonian
\eqref{eq:Ham_rm} has to be solved which requires to perform the exact
diagonalization as described in \secref{ch:diag}.

\begin{figure}[htpb]
  \centering
  \includegraphics[width=1.05\linewidth]{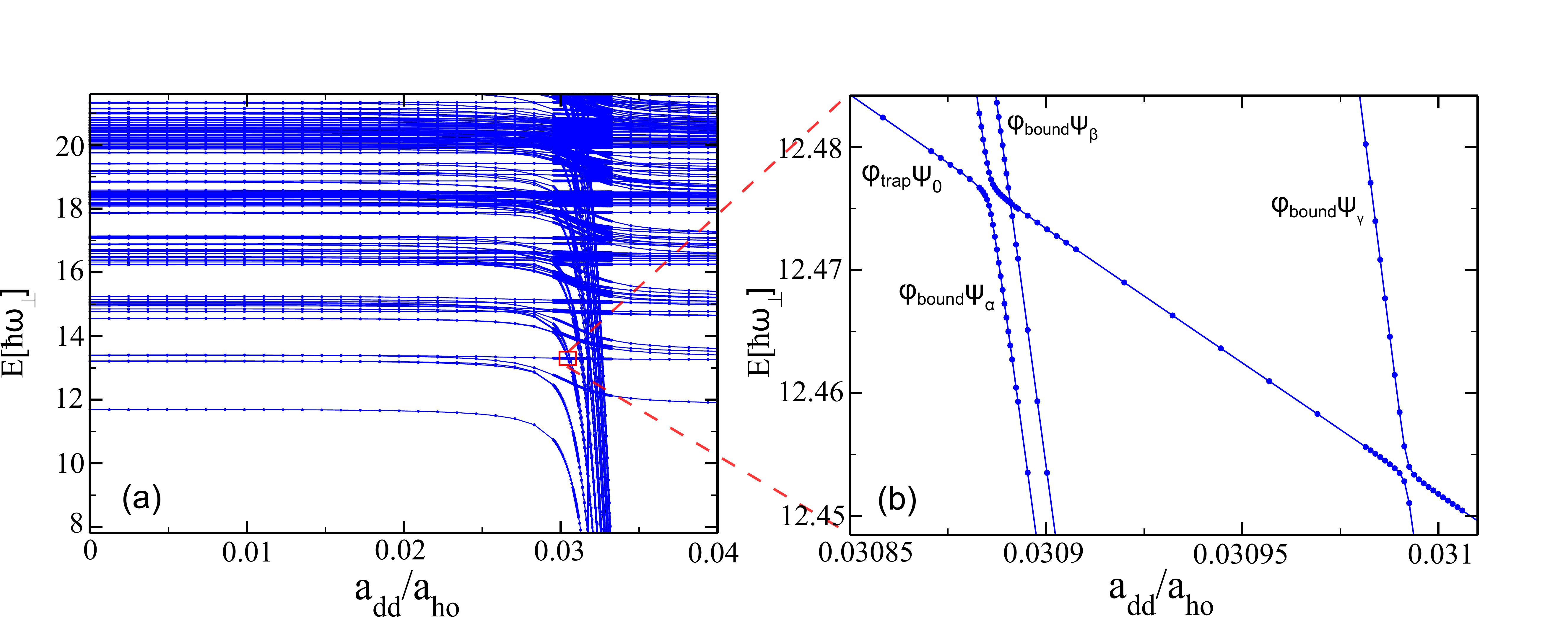}
  \caption{(a) CI spectrum for two aligned dipolar particles for an
    anisotropic sextic trapping potential with the same short-range
    potential as in \figref{Fig:energy_spec}. (b) Magnified view
    of  \figref{fig:CI_Spek} resolving avoided and non-avoided crossings.}
  \label{fig:CI_Spek}
\end{figure}

For distinguishable particles the coupling term $W(\bold{r},\bold{R})$
includes all non-separable parts of the form $r^n R^m$ with $n,m \in
\mathbb{N}\backslash \{0\}$. In the case of identical particles the
non-separable parts consists of monomials with even values of $n$ and
$m$, such as $r^2 R^2$, $r^2 R^4$, and $r^4 R^2$, see
\eqref{eq:sextic_coupling}. The matrix elements with the coupling term
\eqn{W_{\alpha,\beta}=
  \QMa{\Phi^{(\alpha)}(\mathbf{r},\mathbf{R})}{\op{W}(\mathbf{r},
    \mathbf{R})}{\Phi^{(\beta)}(\mathbf{r},\mathbf{R})}\label{eq:coupling_term}}
couple configurations $\Phi$ with \textit{gerade} or \textit{ungerade}
symmetry for even values of $n$ and $m$.  The energy spectra shown in
\figref{fig:CI_Spek} were calculated for two dipolar particles in an
anisotropic sextic trapping potential with the interaction potential
discussed before as a function of the dipole interaction strength
$\frac{a_{\rm dd}}{a_{\rm ho}}$.
\begin{figure}[htpb]
  \centering \includegraphics[width=\linewidth]{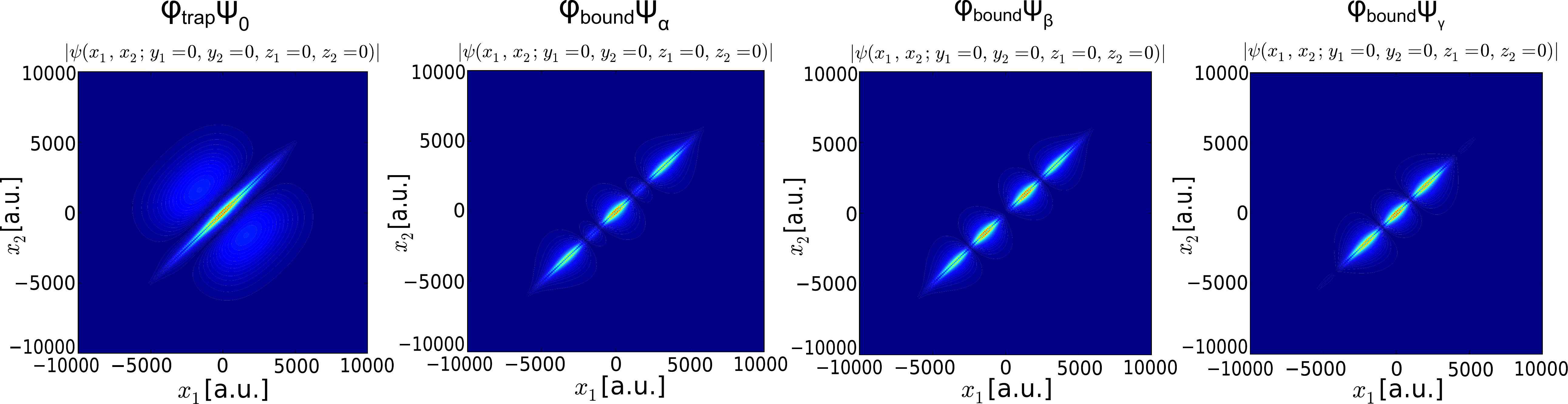}
  \caption{Cuts through the 6-dimensional wavefunctions shown in the
    energy spectrum in \figref{fig:CI_Spek}(b). Cuts for 
    $y_1=0,y_2=0,z_1=0,z_2=0$ are
    displayed. From left to right the cuts, beginning with
    $\varphi_{\rm trap}\;\psi_0$ according to the labelling in
    \figref{fig:CI_Spek}, are shown.}
  \label{fig:WF_CI}
\end{figure}

The major difference between the complete energy spectra for the sextic trap 
compared with the relative-motion spectrum in the harmonic
case is evidently the appearance of the additional center-of-mass 
excitations. While they are present also in the harmonic case, they do 
not couple to the relative motion and can thus be considered separately. 
This is not the case for the sextic trap with coupling. Hence, in the spectrum containing the
center-of-mass excitations (\figref{fig:CI_Spek}) many more states
appear. The configurations of the excited center-of-mass motion bound
states $\varphi_{\rm bound}(\bold{r}) \psi_{0}^{(\rm ex)}(\bold{R})$
cause a very dense area shown in \figref{fig:CI_Spek}(a) for a dipole
interaction strength of about $\frac{a_{\rm dd}}{a_{\rm ho}} \approx
0.03$. By including more configurations this area continues for dipole
interaction strengths above $\frac{a_{\rm dd}}{a_{\rm ho}} \approx
0.03$. However, in the present calculation of the energy spectrum only the relevant
configurations are included, which means only those with small energies 
of a few $\hbar \omega$ are considered. In \figref{fig:CI_Spek}(b) a
magnified view of \figref{fig:CI_Spek}(a) is shown in which avoided
crossings at $\frac{a_{\rm dd}}{a_{\rm ho}} \approx 0.030857$ and
$\frac{a_{\rm dd}}{a_{\rm ho}} \approx 0.030958$ can clearly be
identified. Also a non-avoided (true) crossing next to the first avoided
crossing at $\frac{a_{\rm dd}}{a_{\rm ho}} \approx 0.0309$ is visible.

In the energy spectrum center-of-mass excited bound states cross with
trap states. The anharmonicity in the external potential leads to a
non-vanishing center-of-mass to relative motion coupling. Hence, these
crossings are avoided for certain symmetries of the crossing
states. At the avoided crossing, relative-motion binding energy can be
transferred into center-of-mass excitation energy due to the
anharmonicity in the external confinement.  Since this is an inelastic
process, we denote these resonances as inelastic confinement-induced
dipolar resonances (ICIDR).

These resonances are the dipolar analog for the Feshbach-type
resonances induced by a coupling of center-of-mass to relative motion
systems of ultracold atoms without DDI 
\cite{cold:bold05,cold:pean05,cold:mele09,cold:schn09,cold:kest10,cold:lamp10,cold:vali11}. We
follow the notation introduced in \cite{cold:sala12,cold:sala13} and
denote these resonances as inelastic confinement-induced resonances
(ICIR).  In complete analogy, also in the dipolar gases resonances appear
due to the coupling of center-of-mass excited bound states and trap
states with lower center-of-mass excitation for a non-zero coupling term $W$.
The difference between ICIR and ICIDR is that the adiabatic
transformation of a trap state into a bound state is performed by a
change of the scattering length (for example using a magnetic Feshbach 
resonance) or the geometry of the external
confinement in the case of ICIR and by a variation in dipolar
interaction strength in the case of ICIDR. Moreover, for ultracold
atoms there exists only a single least bound state and the 
scattering length is determined by its position. Hence, for a specific 
center-of-mass
excitation only a single resonance (with the lowest trap state) exists
in the case of ICIR. In contrast, there exists an entire series of
bound states for increasing dipolar interaction strength. Hence, a
complete series of resonances (with the lowest trap state) exist for
each center-of-mass excitation in the case of ICIDR.

To understand the behaviour of the ICIDR in more detail, we investigate the states 
labelled in \figref{fig:CI_Spek}(b). The corresponding
states may be expressed by their 6-dimensional wavefunctions in
absolute coordinates. Representative cuts through the 6D
wavefunctions are shown in \figref{fig:WF_CI}. Since the 
trapping potential is rather anisotropic
$\frac{\omega_z}{\omega_{\bot}} = 10$, the lowest energies correspond
to states which have excitations in the $x$ and $y$ directions, because
the trap is shallower in these directions compared to the tight $z$
direction. In \figref{fig:WF_CI} the wavefunction $\varphi_{\rm
  bound}\psi_{\beta}$ is shown. This state has a node at $(x_1=0,x_2=0)$ and
does not interact with the state $\varphi_{\rm trap}\psi_0$. Hence
the coupling term from equation \eqref{eq:coupling_term} has an
anti-symmetric integrand and therefore the total integral vanishes. 
The vanishing coupling term
\eqn{\QMa{\varphi_{\rm trap}\;\psi_0}{\op{W}(\mathbf{r},
    \mathbf{R})}{\varphi_{\rm bound}\;\psi_{\beta}}= 0 }
leads to a non-avoided crossing between the states $\varphi_{\rm
  trap}\;\psi_0$ and $\varphi_{\rm bound}\;\psi_{\beta}$. Moreover,
the trap state $\varphi_{\rm trap}\psi_0$ and the center-of-mass excited bound
state $\varphi_{\rm bound}\psi_{\alpha}$ have a non-vanishing coupling
term, since these states have the same symmetry, which can be seen
from the nodal structure of the cuts through the wavefunctions in
\figref{fig:WF_CI}. From \figref{fig:WF_CI} it can be concluded that
solely states which have an even (odd) nodal structure can couple to each
other. Otherwise the wavefuntions have different symmetries, i.\,e.\ \textit{gerade} 
and \textit{ungerade}. This results in a vanishing coupling term in
\eqref{eq:coupling_term}, because the total integrand is anti-symmetric. 
It is important to note that it has been demonstrated that the coupling of
trap states to center-of-mass excited bound states at an ICIR has lead to
massive atom losses and heating in a cloud of Cs atoms \cite{cold:hall10b} 
due to molecule formation \cite{cold:sala12,cold:sala13}. As a consequence, 
it is expected that
an ICIDR influences the stability of dipolar quantum gases as well.

\section{Summary and Conclusion\label{ch:summary}}

An approach is presented that allows for the numerical description of
two ultra-cold particles interacting via an arbitrary isotropic
short-range interaction and the long-range, anisotropic DDI confined
to a finite orthorhombic 3D optical lattice. The coupling between
center-of-mass and relative motion coordinates is incorporated in a
configuration-interaction manner and hence the full 6D problem is
solved. A key feature is the use of a realistic inter-particle 
interaction potential, e.\,g.\ numerically provided Born-Oppenheimer
potential curves. The orthorhombic symmetry of the problem, preserved 
by the dipole-dipole interaction, and the quantum statistics
(distinguishable particles as well as identical bosons or fermions) 
are explicitly incorporated in the approach.

With the here presented approach a system of two
dipolar particles interacting via a short-range potential trapped in a
single well of an optical lattice was investigated. It is shown 
that various resonances occur in these systems. The dipole-induced resonances 
occur due to the change of the total interaction potential and
thus the bound-state spectrum as a result of a change of the dipole interaction
strength. Furthermore, the existence of dipolar coupling
resonances of trap states is demonstrated. The position
of specific resonances can be precisely manipulated by a manipulation
of the short-range potential by, e.\,g., a magnetic Feshbach resonance. This
provides additional tools for controlling and manipulating trapped
dipolar quantum gases. This has the potential to provide advanced
cooling schemes, by, e.\,g., performing adiabatic and diabatic changes
of the dipole interaction strength. 

In addition, the occurrence of inelastic confinement-induced dipolar 
resonances (ICIDR) due to a coupling of center-of-mass and relative motion 
is demonstrated. The mechanism of these
resonances is universal and in analogy to inelastic
confinement-induced resonances of ultracold atoms
\cite{cold:sala12,cold:sala13} and Coulomb interacting system like
quantum dots \cite{cold:trop15}. 

As a straightforward extension, the interaction potential of two
non-aligned dipoles is planned. This includes all spherical harmonics
$V_{\rm dd} \propto \sum_{q=-2}^2 Y_2^q $. This leads to a coupling
of both $l$ and $m$ quantum numbers and therefore an even richer 
energy spectrum with much more resonances is expected to be found.

\section*{Acknowledgment}

The authors thank Philipp-Immanuel Schneider for helpful discussions  
and acknowledge financial support from the Humboldt Center for Modern
Optics and the Fonds der Chemischen Industrie. 
One of the authors, S.S., gratefully acknowledges financial support 
from the \textit{Studienstiftung des
deutschen Volkes} (German National Academic Foundation).

\section*{Bibliography}

\bibliographystyle{unsrt} 


\end{document}